\documentclass[10pt, a4paper]{article}
\usepackage{lrec}
\usepackage{multibib}
\newcites{languageresource}{Language Resources}
\usepackage{graphicx}
\usepackage{tabularx}
\usepackage{soul}

\usepackage{epstopdf}
\usepackage[utf8]{inputenc}
\usepackage[english]{babel} 

\usepackage{hyperref}
\usepackage{xstring}

\title{Decentralising power: how we are trying to keep CALLector ethical}

\name{Cathy Chua$^1$, Hanieh Habibi$^2$, Manny Rayner$^2$, Nikos Tsourakis$^2$}

\address{$^1$Independent researcher \\
         $^2$Geneva University \\
         cathyc@pioneerbooks.com.au\\
         \{Hanieh.Habibi,Emmanuel.Rayner,Nikolaos.Tsourakis\}@unige.ch\\}

\abstract{ We present a brief overview of the CALLector project, and consider ethical questions arising from its overall goal of creating a social network to support creation and use of online CALL resources. We argue that these questions are best addressed in a decentralised, pluralistic open source architecture. \\ \newline }

\begin{document}

\maketitleabstract

\section{Introduction}

This paper follows on from another paper presented at the same workshop \cite{ChuaRaynerLeiden2018}, where we present case studies from two large internet communities. Here, we consider the implications for our own project, CALLector. As outlined in the previous paper, the agents we are most worried about controlling are ourselves. If CALLector turns out to be successful, and the opportunity presents itself, experience shows that there is a strong temptation to act unethically.

It is easy to say that we don't need to be concerned about these issues. Few online communities turn into large success stories, and if ours does, we will be delighted. This is lazy and self-deceptive thinking. Basically, it amounts to saying that we would be happy to defraud our members, given the chance, and since it probably will not happen there is no need to worry. The argument does not pass the Kantian test: if everyone thinks this way, then we can be sure that all online communities will exploit and defraud their members. It seems a simple conclusion that our ethical obligation is to develop the network in such a way that we are not in fact planning to exploit our members, even if we are fortunate enough to be given the opportunity. It is worth mentioning explicitly that, at least for some of the authors, the ideas we present here are the fruit of direct and painful experience. Two of us (Chua, Rayner) have been active members of the Goodreads online reviewing community\footnote{\url{https://www.goodreads.com/}. As of April 2019, the site claims to have over 85 million members.} since shortly after its inception. When Goodreads was launched, the founders made many extravagant promises designed to attract new members, in particular that the site would not employ any form of censorship. A few months after the site was sold to Amazon in March 2013, policy abruptly changed, with widespread and arbitrary censorship introduced at zero notice. We were heavily involved in the fightback by the user community and in particular helped create the book which became a focal activity for the protesters \cite{OffTopic}.

In the rest of the paper, we explore the general considerations outlined above in the specific context of the CALLector project. We begin by giving a brief overview of the CALLector project's general goals and the state of play after the first year of development. We list what we see as our main ethical obligations, and outline three plausible ways in which the project could be continued. Finally, we present a brief sketch of the specific solutions we are aiming towards.

\section{Overview of CALLector}
\label{Section:CALLector}

CALLector \footnote{\url{https://www.unige.ch/callector/}} is a project funded by the Swiss National Science Foundation and based at Geneva University; it offically started on April 1 2018, and is scheduled to run until December 31 2021. Its overall goal is to create a social network designed to support users who wish to create, use and share online CALL content. The key word here is ``create'': we want it to be possible for users to create their own content, using suitable tools. The abstract question we wish to investigate is whether the well-documented rewards associated with working within a social network --- basically, various kinds of positive social feedback from the other members --- can motivate people to create large amounts of useful content. This model has worked well for online communities centred around, for example, book reviews (Goodreads), cartographic data (OpenStreetMap\footnote{\url{https://www.openstreetmap.org}}) and knitting patterns (Ravelry\footnote{\url{https://www.ravelry.com}}), so maybe it will also work for CALL. 

\begin{figure*}[!h]
\begin{center}
\includegraphics[width=11cm]{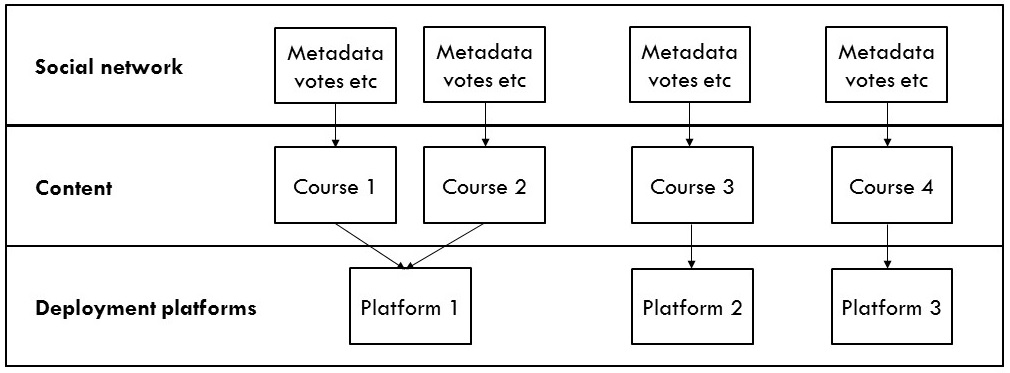} 
\caption{Planned CALLector architecture}
\label{figure:CALLectorAbstract}
\end{center}
\end{figure*}

\begin{figure*}[!h]
\begin{center}
\includegraphics[width=11cm]{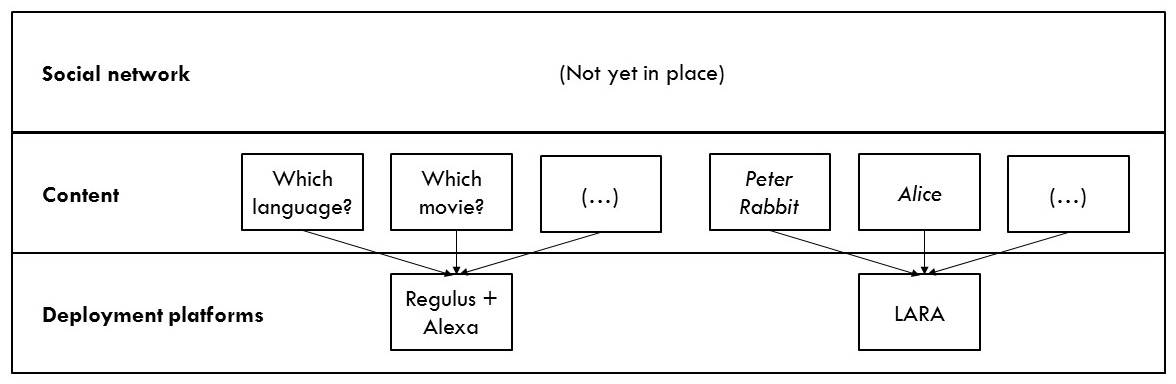} 
\caption{Current CALLector architecture}
\label{figure:CALLectorCurrent}
\end{center}
\end{figure*}

The global architecture is shown in Figure~\ref{figure:CALLectorAbstract}, and consists of three layers. In the middle, we have the content. Underneath, we have the platforms that deploy the content. At the top, we have the social network, which indexes the content and allows users to perform functions like rating, commenting, recommending and so on. Figure~\ref{figure:CALLectorCurrent} shows the current state of instantiation of this architecture. The social network level does not yet exist: work is just starting now in April 2019. We have two deployment platforms, Regulus/Alexa and LARA, and some initial content. Work to date on the Regulus/Alexa platform is described in our other companion paper in these proceedings \cite{TsourakisEALeiden2018}. Very briefly, the platform allows construction of interactive CALL games that can be deployed on Amazon Alexa devices like the Echo. Games are written in a spreadsheet form similar to Alexa's ``blueprints''\footnote{\url{https://blueprints.amazon.com}}, but oriented towards the requirements of CALL. We have so far constructed about twenty sample games, and have a couple of external content-creators whom we are informally supporting. 

\begin{figure*}[!h]
\begin{center}
\includegraphics[width=14cm]{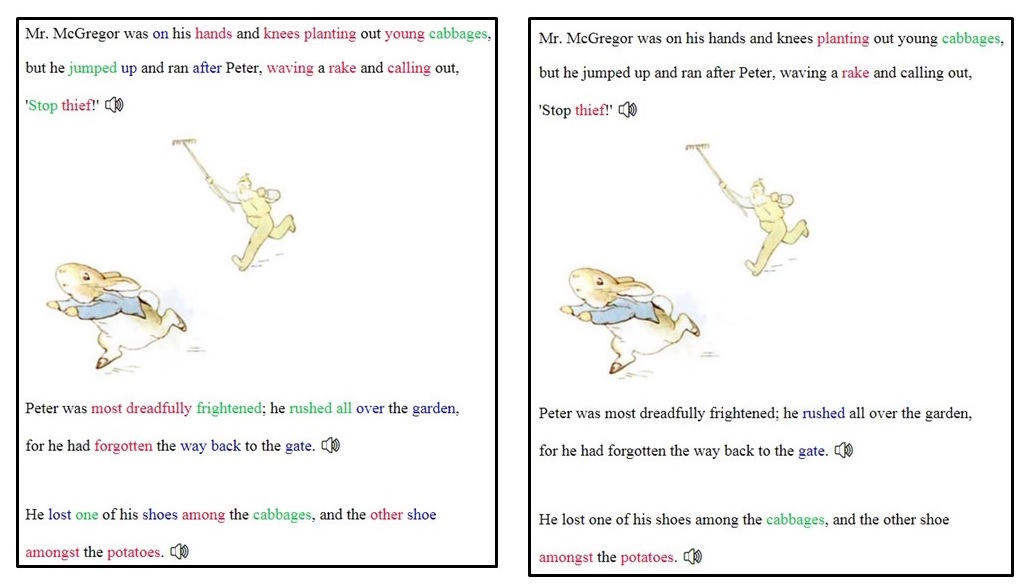} 
\caption{Example of text shown by LARA: two paragraphs from \textit{Peter Rabbit} marked up (left) after completing \textit{Peter Rabbit} but nothing else, and (right) after reading both \textit{Peter Rabbit} and the first three chapters of \textit{Alice in Wonderland}.}
\label{figure:LARA}
\end{center}
\end{figure*}

\begin{figure*}[!h]
\begin{center}
\includegraphics[width=14cm]{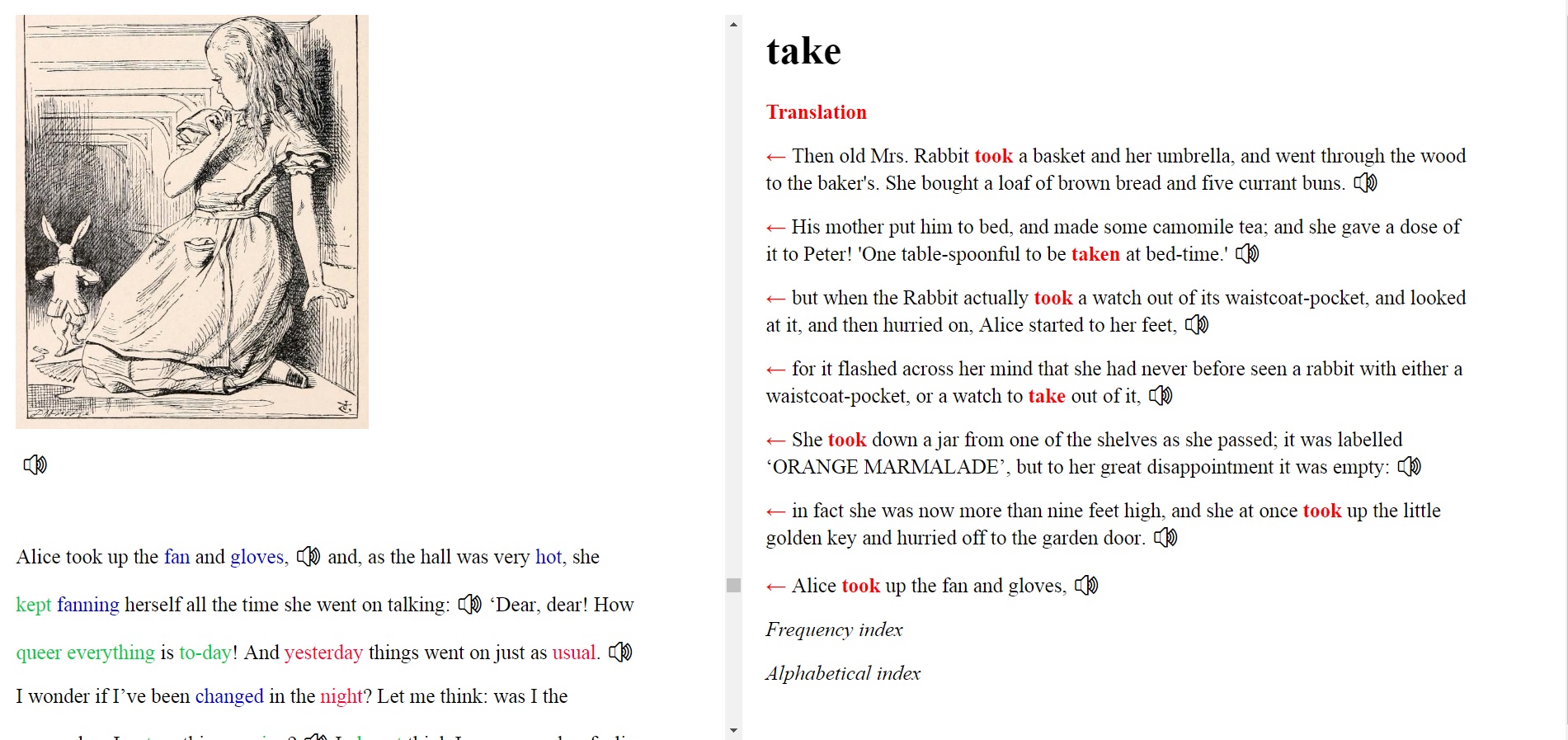} 
\caption{Example of content produced by the current LARA prototype, showing a personalised reading progress. The student has so far read \textit{Peter Rabbit} followed by the first three chapters of \textit{Alice in Wonderland}. The left hand side shows the marked-up text, where the student has just clicked on the word ``took''. The right hand side displays occurrences of different inflected forms of ``take'' in both source texts.}
\label{Figure:PeterAlice}
\end{center}
\end{figure*}

LARA (Learning And Reading Assistant), which will be described at greater length elsewhere, is a platform we started developing in Q3 2018\footnote{The state of play of the project as of late February 2019 is summarised in our informal position paper \cite{LARAPositionPaper}. Examples of LARA content are posted at \url{https://www.unige.ch/callector/lara-content/}.}. As the name suggests, the goal is to provide assistance to students who are improving their competence in the L2 through reading. LARA processes text into a marked-up form which offers two basic kinds of support. First, each sentence is linked to a recorded audio file, so the student can always find out what the text they are reading sounds like. Second, and more unusually, there is a personalised hyperlinked concordance which shows the student where each word has previously occurred \textit{in their own reading experience}. Figure~\ref{figure:LARA} illustrates LARA's core functionality using a short passage from \textit{The Tale of Peter Rabbit}. The marked-up version of \textit{Peter} is shown at two points in the learner's reading progress: on the left, where the learner has read the whole text and nothing else, and on the right, where they have read both \textit{Peter Rabbit} and the first three chapters of \textit{Alice in Wonderland}\footnote{The two version can be found online at \url{https://www.issco.unige.ch/en/research/projects/peter_rabbitvocabpages/_hyperlinked_text_.html} and \url{https://www.issco.unige.ch/en/research/projects/callector/reader1_englishvocabpages/_hyperlinked_text_.html.}}. Colours are used to indicate how many times each lemma occurs in the text; words in black have occurred more than five times, words in red only once, while blue and green show intermediate values. As the picture shows, the colours effectively track the learner's increased exposure to vocabulary between the two snapshots. In the first snapshot, only function words and a few content words central to the story appear in black; in the second, many of the words marked in red have turned black or blue, indicating that they have been read several times during the intervening period. The text has been manually divided into segments and recorded in audio form. A loudspeaker icon marks the end of each segment, and the learner can listen to the segment in question by clicking on the icon. The learner can click on any word and get a personalised concordance page containing up to ten segments where the word appears (Figure~\ref{Figure:PeterAlice}).

LARA is designed on the assumption that content will in general be distributed over multiple servers on the web, with a master file that associates resource identifiers with URLs. When constructing the personalised concordance pages, the compiler downloads as little as possible, leaving the bulky multimedia files on the remote servers and inserting links to them where necessary. There is thus no need to have all the user data on one server either.

\section{Ethical issues}

We now move on to ethical issues; we begin by listing the people and organisations to whom we have obligations. We consider these to be i) ourselves, ii) the project funder (the Swiss National Science Foundation), iii) our employer (Geneva University), iv) the network's content creators (under ``content creators'', we also include external people who contribute to the system-level architecture), and v) the network's content users. In our capacity as the people responsible for carrying out the project, we consider our main obligations to these agents to be roughly the following.

\begin{description}

\item[Ourselves:] Keep our jobs. Publish. One of us needs to complete a PhD.

\item[Funder:] Carry out the project as described in the proposal. Publish.

\item[Employer:] Publish. Attempt to leverage the results of the project to bring in more funding. 

\item[Content creators:] Provide stable, maintainable hosting for content. Provide stable, maintainable hosting for user data. Respect the content creators' IP rights. Respect the content creators' rights as members of the social network.

\item[Content users:] Provide stable, maintainable hosting for content. Provide stable, maintainable hosting for user data. Respect the content users' rights as members of the social network.

\end{description}

Based on the above, we consider three generic strategies for continuing the project, and to what extent they will fulfil our ethical obligations.

\subsection{Default academic project}

If we take no special steps, we expect the project to develop roughly as follows. We will keep on haphazardly extending the system in order to support short-term activities, primarily writing papers, getting a PhD, and perhaps doing data collection. The software base will consist of messy research code, each part of which is typically understood by only one person. There will be minimal documentation. A year or two after funding ends, the platform will stop working.

This outcome would minimally fulfil our obligations to ourselves and the funder. If the result was good enough that we were able to submit a proposal for some kind of follow-on project, it would minimally fulfil our obligations to our employer. It would however leave the content creators feeling betrayed and the content users disappointed.

\subsection{Aim towards commercialisation}

A second strategy would aim towards commercialising the project. This would require more careful extension of the system, done in a way that prioritised creation of a substantial and growing social network; the worth of the project would mostly depend on the size of the network. Critical activities would be to develop easily usable tools for content creation, gamify the site to make usage addictive, and ensure that hosting was scalable and reliable. Code would need to be more carefully maintained and documented, and at some point would be moved into a private repository. The long-term goal would be to sell the project to whoever was willing to buy it, most likely a large multinational.

If this strategy succeeded, we would strongly fulfil our obligations to ourselves, the funder, and our employers. Content creators would however again feel betrayed; their unpaid work would have been exploited to make money for us and/or our employer. Whether content users considered that we had fulfilled our obligations to them would depend on the strategy followed by the project's new owners. 

\subsection{Aim towards viable open source project}

A third strategy would aim towards creation of a viable open source project. Goals would overlap to some extent with those in the commercialisation scenario: in particular, central activities would again be to develop tools for content creation and to ensure stable hosting. Code would however be kept open source, and there would be a strong emphasis on long-term maintainability. This would require more extensive documentation and code reviewing, in particular ensuring that multiple people were involved in maintaining each module.

A success here would strongly fulfil our obligations to ourselves, the funder, content creators and content users. It would at least weakly fulfil our obligations to our employer. 

\section{Options for creating an ethical project}

The discussion in the preceding section and the case studies considered in the linked paper suggest that an open source strategy is strongly preferable on ethical grounds: it is the only one which respects the rights of the content creators and content users, who are essential to the success of the project. We now go on to consider how such an open source version of CALLector might be realised. We divide up material under four main headings; decentralisation, ownership, third-party resources and long-term maintainability.

\subsection{Decentralisation}
\label{Label:Decentralisation}

In many online communities, centralised power is the core tool used to exploit the members. The community's founders maintain the servers on which the community resides. New members sign, usually without reading it, an EULA which gives the founders more or less absolute power over their activities. In particular, members can always be blocked or thrown out, and if they do they have no recourse.

We propose the following remedy to this problem. Rather than requiring that all the activities of the social network be carried out on the founders' dedicated servers, we are instead developing tools which offer the option of being used on machines controlled by other members of the community. We consider specific issues concerned with \textit{developing}, \textit{deploying} and \textit{discussing} content.

\begin{enumerate}
\item The development tools are used to author CALL courses and compile them into deployable content. These tools are open source, and configurable to run both on local machines and on a remote server. Users who do not wish to be bothered with the inconvenience of installing the tools will be able to run them on our servers; users who prioritise independence will be able to download them and run them locally.

\item Issues concerning deployment differ sharply between the two component platforms. LARA content, which takes the form of ordinary web pages, is unproblematic and can be straightforwardly uploaded to a webserver. The nontrivial issues arise with regard to the interactive speech content produced by Regulus. We discuss these below in \S\ref{Section:ThirdParty}

\item As far as \textit{discussing} content goes (the social network part of the project), concrete development has not yet started. Our plan is to use a peer-to-peer architecture where information is not stored on a single server, but distributed across multiple servers \cite{yeung2009decentralization}. Methods of this kind are explicitly designed to address the issues we discuss here.

\end{enumerate}

\subsection{Ownership}

Many online communities make it a practice to require that the community founders gain full non-exclusive ownership of user-created content, in exchange for hosting said content. Since we are explicitly aiming not to place members under the obligation to rely on us to host their content, we conversely do not aim to acquire rights to this content, which is likely to place us in an ethically dangerous position. We will, instead, take the default view that a content creator's content belongs to the creator and no one else. The model is that we are providing compilers and other development tools: a compiler supplier does not normally claim ownership of anything produced using their compiler. 

We will however encourage content creators to release their content in open source form, under standard open source licenses like LGPL. The basic encouragement we offer is altruistic: if you make your content open source, other people can use and adapt it, and that makes you a more valued member of the community. (Pure altruism may well be reinforced by some kind of credit/kudos mechanism at the social network level). Experience shows that many people find these kinds of reward motivating, and in practice we can expect many people to want to take part as open source developers.

\subsection{Third-party dependencies}
\label{Section:ThirdParty}

The content produced by LARA consists of ordinary web pages, and third-party dependencies are limited to those inevitable when using the internet in any form. For the speech-enabled content produced by Regulus, the issues are more complex, and different content creators will have different priorities. Since our whole enterprise is founded on the goodwill of the content creators, it seems to us that we should try to accommodate as broad a spectrum of content creator profiles as is feasible.

Based on preliminary discussions, we can see at least two  types emerging. The first class is pragmatic in focus. Their main interest is in getting the content up and running reliably with as small an investment of effort as possible, and they also want to retain the option of monetising the content if demand reaches the point where this is feasible. For this kind of content creator, Amazon Alexa is an attractive deployment vehicle. Alexa offers high-quality hands-free far-field recognition on an automatically scalable platform. About a hundred million Alexa-enabled devices have already been sold, and Alexa support will soon be available on many new laptops, so an Alexa-deployed course will reach a wide audience. Alexa also makes it straightforward to monetise apps.

We also see a class of content creators who place a higher priority on ethical issues; some of these people have reservations about Amazon's ethical policies, and  have expressed unhappiness about the idea of using Amazon products. Partly with this class of user in mind, and partly on the general principle that it is unwise to be reliant on a single third-party supplier, we are developing an alternate deployment platform for Regulus-derived content. This consists of a Django wrapper for the core runtime code which allows it to be run as a web service, together with a client which performs speech recognition using the Google Speech API. The problem, of course, is that the content creator is still reliant on a third-party supplier, which will now be Google rather than Amazon; unfortunately, it is extremely challenging to create open source cloud-based recognition resources which can realistically compete with the ones offered by these large multinationals. But our architecture will at least make it easy to integrate other third-party recognisers as they emerge.

\subsection{Long-term maintainability}

It is obviously not possible to us at this stage to present a plan which guarantees the planned CALLector community's long-term maintainability. It is ethically reasonable, however, to demand that we provide evidence that the community has good prospects for being maintained if it grows to a size where there is a large enough user base.

Again, we see a clear difference between the two component platforms. LARA's codebase is small and simple enough that we expect many people could potentially maintain it. In addition, the distributed design means that it does not require a large server farm to operate, but can comfortably be spread over multiple independent servers, none of which would need to bear a heavy load. It seems to us quite reasonable that a usable free network could be run by a loose federation of server operators, perhaps mostly academic. Someone who wanted access to LARA would only need find a convenient server near them that was prepared to host their data.

Some aspects of this proposed organisation carry over to the Regulus platform, but the same problem arises: until free open source speech recognition resources are available, the community requires access to a third-party commercial agent in order to function. The preconditions for long-term maintainability are thus less clear.

\section{Other ethical issues}

The focus of this paper has been on ethical issues having to do with our obligations to the various stakeholders in CALLector, but for completeness we briefly outline our position on two other topics related to the project: privacy and copyright.

\subsection{Privacy}

Privacy issues are important for the LARA platform, and enter in two ways. First, since the unique feature of LARA is precisely that it gives the user a text marked up on the basis of their own reading experience, it is \textit{a fortiori} necessary for the platform to store that reading experience in some form. We are mindful of the fact that this is potentially sensitive information. At the moment, it is hard to believe that anyone will care if someone else discovers that they have read \textit{The Tale of Peter Rabbit} or \textit{Alice in Wonderland}. Since the intention is to make it possible for content creators to upload a wide variety of texts, the situation may however change in the future. For students living in countries which operate religious or politically motivated censorship policies, it is certainly conceivable that their online reading list may be information they need to keep private. Our initial plan here is the obvious one: we will store information about reading histories on secure servers on the university cloud, we will not require users to give us any information which might be used to identify them, and we will provide access over HTTPS. Later in the project, we will make available the means to allow people to set up their own servers, which perform relevant processing --- constructing sets of concordance pages, etc --- on the local machine, keeping the reading history there as well; this will make it possible to parties who do not trust our security to run on machines under their physical control. The distributed nature of the LARA architecture (cf. remarks at the end of \S\ref{Section:CALLector}) means that personal servers can be made easy to install and run.

A second and less central set of privacy issues is related to  logging of user activities in LARA. The current prototype uses static web pages, but we plan at a later stage to move to dynamic page generation. The primary motivation is efficiency, but this will also make it feasible to perform fine-grained logging of user activities at the level of accessing concordance pages, playing audio files and looking up translations. This kind of data could be of considerable scientific interest, but initial polling suggests to us that many users will be unhappy about such detailed recording of their actions. We will consequently only do this when users have clearly given informed consent.

We should add that we have been criticised on ethical/privacy grounds for encouraging children to use the Alexa CALL games we have developed. We do not think this criticism is well founded. First, millions of children all over the world are already using Alexa games which from the privacy point of view are similar to ours, and the number is rapidly increasing. Our decisions will have no influence on this trend. Second, and even more to the point, we are not aware of serious evidence that suggests Alexa is invading user privacy more than a multitude of other web and cellphone technologies. We have discussed the issue with Amazon engineers, who have convinced us that Amazon is telling the plain truth when they say that Alexa's listening capability is only switched on after speaking the ``wake word''. The thing that makes their argument so plausible is that Alexa's performance would in fact be a great deal better if it listened all the time. At the moment, the problem is rather in the opposite direction: in order to reduce server bandwidth, Alexa drops out of the current app if the user fails to respond within a few seconds. This behaviour causes many problems at the user interaction level, and Amazon do not even provide a toggle that allows it to be switched off.

\subsection{Copyright and IPR}

Since LARA makes integral use of published texts, copyright issues naturally arise. Our position here is simple. As previously mentioned, we consider the LARA tools to have the status of open source compilers which we make available to the community. We allow anyone to use the tools as they wish, claim no intellectual property rights with regard to LARA documents which users may produce, and take no responsibility for any possible consequences. In particular, a user who processes text and posts it in LARA form takes responsibility for any relevant copyright issues. We will not take responsibility for checking the copyright status of LARA texts that may be posted on our own servers, but rather use the common social network approach of making it easy to flag potentially offending texts. If texts are flagged as infringing copyright or being offensive (hate speech, pornography), we reserve the right to remove them. We will in normal cases warn the content creator in advance, giving them a grace period to correct or save the content.

\section{Summary and conclusions}

We have presented a brief overview of CALLector and outlined how we intend to progress it, focusing in particular on ethical aspects. There is a substantial difference between the two underlying platforms, LARA and Regulus. For LARA, we see a direct path towards an ethically attractive solution. The open source compiler is small and simple, and could feasibly be maintained by people other than ourselves. It is easy to use, and the content is trivial to deploy. The situation is far less clear with Regulus, where it seems right now difficult to deploy content without incurring a critical dependence on a large multinational, either Amazon or Google. Of course, speech-enabled content is becoming increasingly important and speech-enabled CALL software is potentially very useful. We do not think it is right for us to make ethical decisions on behalf of our potential users, and intend to move forward on making both platforms available. 

We note that so far we have seen much more interest in LARA. An interesting recent development, however, is that several users have asked us whether it would be feasible to merge the platforms, optionally providing speech recognition capabilities inside the LARA environment that could give students feedback when reading LARA text aloud. This opens up a whole new set of issues, which we look forward to exploring.


\section{Acknowledgements}

We would like to thank Branislav Bédi and Karën Fort for organising the enetCollect workshop at which this work was originally presented and for many useful and stimulating discussions. Johanna Gerlach has been extremely helpful in supporting the LiteDevTools platform, which is heavily used by both the Regulus/Alexa and LARA platforms. Many people have now been involved in developing content. We would particularly like to recognise the contributions made by Elham Akhlaghi (Farsi), Branislav Bédi (Icelandic), Matt Butterweck (German and Middle High German; also code development), Monica Depasquale (Latin), Pierre-Emmanuel Gallais (French), Junta Ikeda (Japanese), Annabel Keigwin (French and English) and Sabina Sestigiani (Italian).

\section{Bibliographical References}

\bibliographystyle{lrec}
\bibliography{proposal}

\begin{thebibliography}{}

\bibitem[\protect\citename{Akhlaghi \bgroup et al.\egroup
  }2019]{LARAPositionPaper}
Akhlaghi, E., Bedi, B., Chua, C., Habibi, H., and Rayner, M.
\newblock (2019).
\newblock {LARA}: A learning and reading assistant.
\newblock Position paper.
  \url{https://www.issco.unige.ch/en/research/projects/callector/LARA_position_paper.pdf}.

\bibitem[\protect\citename{Chua and Rayner}2019]{ChuaRaynerLeiden2018}
Chua, C. and Rayner, M.
\newblock (2019).
\newblock What do the founders of online communities owe to their users?
\newblock In {\em Proceedings of the enetCollect WG3/WG5 workshop}, Leiden,
  Holland.

\bibitem[\protect\citename{Reader}2013]{OffTopic}
Reader, G.
\newblock (2013).
\newblock {\em Off-Topic: The Story of an Internet Revolt}.
\newblock \url{https://www.goodreads.com/ebooks/download/18749172-off-topic}.

\bibitem[\protect\citename{Tsourakis \bgroup et al.\egroup
  }2019]{TsourakisEALeiden2018}
Tsourakis, N., Rayner, M., Gallais, P.-E., Habibi, H., Chua, C., and
  Butterweck, M.
\newblock (2019).
\newblock Alexa as a {CALL} platform for children: where do we start?
\newblock In {\em Proceedings of the enetCollect WG3/WG5 workshop}, Leiden,
  Holland.

\bibitem[\protect\citename{Yeung \bgroup et al.\egroup
  }2009]{yeung2009decentralization}
Yeung, C.-m.~A., Liccardi, I., Lu, K., Seneviratne, O., and Berners-Lee, T.
\newblock (2009).
\newblock Decentralization: The future of online social networking.
\newblock In {\em W3C Workshop on the Future of Social Networking Position
  Papers}, volume~2, pages 2--7.

\end{thebibliography}

\end{document}